# A ROOM TEMPERATURE POLARITON LIGHT-EMITTING DIODE BASED ON MONOLAYER $WS_2$

Jie Gu,[†,‡,¶] Biswanath Chakraborty,[†,¶] Mandeep Khatoniar,[†,‡] and Vinod M. Menon[*,†,‡]

†*Department of Physics, City College of New York, City University of New York,*

‡*Department of Physics, The Graduate Center, City University of New York,*

¶*Contributed equally to this work*

*E-mail: vmenon@ccny.cuny.edu

**Half-light half-matter quasiparticles termed exciton-polaritons arise through the strong coupling of excitons and cavity photons. They have been used to demonstrate a wide array of fundamental phenomena and potential applications ranging from Bose-Einstein like condensation[1–3] to analog Hamiltonian simulators[4,5] and chip-scale interferometers[6]. Recently the two dimensional transition metal dichalcogenides (TMDs) owing to their large exciton binding energies, oscillator strength and valley degree of freedom have emerged as a very attractive platform to realize exciton-polaritons at elevated temperatures[7]. Achieving electrical injection of polaritons is attractive both as a precursor to realizing electrically driven polariton lasers[8,9] as well as for high speed light-emitting diodes (LED) for communication systems[10]. Here we demonstrate an electrically driven polariton LED operating at room temperature using monolayer tungsten disulphide ($WS_2$) as the emissive material. To realize this device, the monolayer $WS_2$ is sandwiched between thin hexagonal boron nitride (hBN) tunnel barriers with graphene layers acting as the electrodes[11,12]. The entire tunnel LED structure is embedded inside a one-dimensional distributed Bragg reflector (DBR) based microcavity structure. The extracted external quantum efficiency is ~0.1% and is comparable to recent demonstrations of bulk organic[13] and carbon nanotube based polariton electroluminescence (EL) devices[14]. The possibility to realize electrically driven polariton LEDs in atomically thin semiconductors at room temperature presents a promising step towards achieving an inversionless electrically driven laser in these systems as well as for ultrafast microcavity LEDs using van der Waals materials.**

Atomically thin van der Waals (vdW) materials have become a very attractive platform for realizing plethora of fundamental phenomena and technological innovations owing to their highly desirable electrical, optical, mechanical and thermal properties. Of these vdW materials, TMDs have become extremely attractive for optoelectronics owing to their unprecedented strength of interaction with light. Combined with other vdW materials such as graphene, a conductor and hexagonal boron nitride (hBN), an insulator, one can realize the entire gamut of electrically driven semiconductor devices such as LEDs, photodetectors, sensors, and energy storage devices[15]. Owing to their large exciton binding energy in the monolayer limit combined with the properties such as valley polarization, the TMDs have also become a highly sought after platform for realizing strongly coupled exciton-polariton devices with largely unexplored characteristics such as the valley degree of freedom, charged excitons, long distance propagation and excited states[16–21]. Most of the work on exciton-polaritons based on two-dimensional (2D) TMDs have been done via optical excitation as has been the scenario for most of the field of exciton-polaritons. However with the recent emergence of polaritonic devices for applications ranging from ultrafast LEDs to polaritonic circuits[22,23], there is much interest in realizing electrically driven polariton emitters. Such emitters are highly desirable and also markedly distinct from their optically driven counterparts due to the device complexity. Polariton LEDs have been demonstrated in traditional inorganic semiconductors[8,9,14,24–28] as well as in organic materials[13,29–31] using bulk materials or with multiple quantum wells. While there have been few reports of control of strong coupling in 2D TMDs via electric field gating[32,33], there has yet to be any demonstration of electrical injection of exciton-polaritons and electroluminescence (EL) from such strongly coupled systems. Here we demonstrate an electrically driven polariton LED using vdW heterostructures that operates in a tunnel injection architecture where the excitons in the TMD monolayer is strongly coupled to cavity photons. The attractiveness of the 2D material platform stems from the possibility to realize devices that have atomically thin emissive layers which can be integrated with other vdW materials for contacts (graphene) and tunnel barriers (hBN). Furthermore, the 2D material platform also presents the unique opportunity to integrate these polariton LEDs with other vdW materials with magnetic[34], superconducting[35] and topological transport properties[36] resulting in hitherto uncharted device features.

Shown in Fig. 1a is the schematic of the device. There are silver and twelve periods of distribute Bragg reflector (DBR) acting as the cavity top and bottom mirror, respectively. In the active region, we have the tunnel area as well as two more hBN encapsulated monolayer $WS_2$. The tunnel area consists of a vdW heterostructure with monolayer $WS_2$ as the light emitter, thin layers of hBN on either side of monolayer acting as the tunnel barrier and graphene as transparent electrodes to inject electrons and holes. The two more $WS_2$ layers are included to increase the overall oscillator strength and thereby result in pronounced Rabi splitting of the polariton states. The cavity mode is tuned by the thickness of PMMA. Details of the sample preparation and optical response of the empty cavity are discussed in the Methods section and the Supplementary Section S1, respectively. Figure 1b shows the optical microscope image of the vdW heterostructure on the bottom DBR. Due to the high reflectivity of the bottom DBR, the tunnel region in Fig. 1b has a very low reflection contrast, resulting with only top thick hBN layer observable. Further images of the device at various stacking steps of the van der Waals heterostructure are shown in Supplementary Section S2.

The band diagram of the vdW heterostructure in the tunnel geometry under bias is shown in Fig. 1c. Electroluminescence (EL) is observed above the threshold voltage when the Fermi level of top (bottom) graphene is biased above (below) the conduction (valence) band of $WS_{2,}$ allowing electron (holes) to tunnel into the $WS_2$ conduction (valence) band. This creates favorable condition for exciton formation within the $WS_2$ layer, followed by the electron-hole radiative recombination. Unlike p-n junction based light emitters, which rely on doping for operation[37–39], EL from the tunneling devices solely rely on the tunneling current, thus avoiding optical losses and any variation of resistivity with temperature. At the same time, the tunnel architecture allows much larger emission region as compared to p-n junction based TMD devices. Figure 1d shows the electrical characteristics of tunneling current density J as a function of bias voltage V between the graphene electrodes. The sharp rise in current for both positive and negative voltages indicates the onset of tunneling current through the structure. With an optimum thickness of hBN layers (~2 nm), we ensured to observe significant tunnel current and increased lifetime of injected carriers for radiative recombination.

Before we perform the EL experiment, we characterize our device by angle resolved white light reflectivity and PL to ascertain that we are indeed in the strong coupling regime. These measurements as well as the EL are carried out using a Fourier space (k-space) imaging set up to map out the energy versus in-plane momentum dispersion. Figure 2a shows the angle resolved reflection spectra from the active area of our device demonstrating an anti-crossing behavior. We further observed photoluminescence (PL) under non-resonant excitation (460 nm) showing an intense emission from the lower polariton branch and weaker emission from the upper polariton branch as shown in Fig. 2b (see Methods section for optical measurement details). The EL measurements are carried out under an external dc bias applied using a Keithley 2400 source meter. The angle resolved dispersion of polariton EL at 0.1 $\mu A/\mu m^2$ injection is shown in Fig. 2c and is found to be identical to the PL dispersion (Fig. 2b). Spatial image of the EL is shown in Supplementary Section S3. The Rabi splitting, and cavity detuning derived using coupled oscillator mode fit (shown by solid and dashed white lines) to the EL experiment is $\sim$ 33 meV and $\sim$ -13 meV, respectively. The cavity detuning is defined by $\delta = E_c - E_x$, where $E_x$ is the exciton energy and $E_c$ stands for cavity photon energy with zero in-plane momentum. The sectional slice, at different angles, from the EL dispersion is shown in Fig. 2d. The dispersion of the upper and lower polariton modes can be clearly seen here with the anticrossing occurring in the vicinity of the exciton resonance (solid line).

As the tunneling current is increased, the overall intensity of EL goes up. Weak EL from the polaritons is observed near threshold bias (Fig. 3a), while at sufficiently higher bias above the threshold, the polaritonic emission becomes distinctively bright (Fig. 3b). Shown in Fig. 3c is the polar plot of EL intensity as a function of angle depicting a narrow emission cone of ±15$^{o}$. The radiation pattern remaining almost unchanged for both minimum (green curve) and maximum (orange curve) driving current. The integrated intensity under different driving tunnel currents is shown in Fig. 3d (black dot, left axis) and follows an almost linear trend. Increasing current to sufficiently higher values could lead to successful polariton scattering along the lower branch and create extremely narrow emission pattern due to polariton lasing. However, in our case we were limited by the dielectric breakdown of hBN tunneling barrier and hence could not reach this regime. Improvement in the quality of hBN could further increase the damage threshold. The external quantum efficiency (EQE) which is the ratio of the number of extracted photons to the number of injected charge particles, is also plotted in Fig. 3d (red dots, right axis) as a function of current density. The observed EQE is comparable to other reports of polariton LEDs such as in organic materials [13] and carbon nanotubes[14], albeit the light emitting layer of the present device is only few atom layer thick (~0.7nm) compared to the much thicker active material used in previous demonstrations. It should however be noted that the observed EQE is lower than that reported for similar tunneling devices not confined a cavity geometry[11,40]. The reduced efficiency is likely due to the poor light extraction from our cavity, which needs further improvement as well as in-plane waveguiding. An alternative way to increase EQE is to stack more monolayers inside the tunnel region separated by thin hBN[11]. Details of the EQE estimation is given in Supplementary Information Section S5.

We also investigated the effect of the cavity detuning on the polariton EL by fabricating a similar device but with a larger cavity detuning (- 43 meV). Further details of this device (Device 2) are discussed in Supplementary Section S4. Figure 4a shows the angle resolved EL spectra from the highly negatively detuned device at a current density of 0.2 $\mu$A/$\mu$m$^2$. Owing to the larger detuning, this device shows a strong bottle neck effect in the EL with the emission maximum occurring at a large angle. This is further confirmed in Fig. 4b which compares the normalized polar plot from Device 2 with that obtained from Device 1 (Fig. 2c). For the higher negative detuning, emission maximum occurs at 18 degrees (blue curve) as compared with device 1 (orange curve) which centers at 0 degree.  This bottleneck effect for the larger detuning sample can be understood as a result of poor polariton scattering to $k_{||} = 0$ owing to the short polariton lifetime in these cavities.

In summary, we have demonstrated room temperature polariton EL from a vdW heterostructure embedded in a microcavity. The tunneling architecture of our device enables electron/hole injection and recombination in $WS_2$ monolayer, which acts as the light emitting layer. The tunneling mechanism of the device does not require any doping of the constituents, thus minimizing losses and temperature related variations. The entire tunnel LED comprising of few layer graphene contacts, hBN tunnel barriers and encapsulating layers are embedded in a microcavity and the strong coupling regime is achieved as indicated by the presence of the two polariton branches in reflectivity, PL and EL. Above certain threshold bias, the bands are aligned and favors carrier tunneling from graphene electrodes to the monolayer $WS_2$ through the ultrathin hBN barriers. Varying current injection above the threshold leads to significant increase in emission intensity. The EL is also found to be highly directional owing to the cavity dispersion. Further improvement in cavity Q factor and higher current injection should help realize more efficient microcavity LEDs and the possibility of an electrically driven low-threshold microcavity polariton and/or a photon laser. The present demonstration of EL from TMD exciton-polaritons in a microcavity is a significant progress towards realizing such electrically driven integrated microcavity light emitters using 2D vdW materials for potential application as ultrafast LEDs and low threshold lasers.

**Methods:**

***Sample Preparation.*** The DBR consisting of 12 periods of alternate layers of $SiO_2$ (106.2 nm) and $Si_3N_4$ (77.5 nm) was grown on silicon substrate by plasma enhanced chemical vapor deposition (PECVD) using a combination of nitrous oxide, silane and ammonia under a temperature of 350°C. Two gold contacts were then prefabricated onto the DBR top surface. We used electron beam lithography to write the contacts pattern and deposited Ti/Au (2nm/8nm) by electron beam evaporation. Monolayer $WS_2$, graphene and multilayer hBN were exfoliated from bulk crystals (WS2 and graphene from HQ Graphene and hBN from 2Dsemiconductor Inc.) using scotch tape f onto 300nm $SiO_2$/Si substrate. Heterostructure stacking and transfer were done using the well-known poly-propylene carbonate (PPC) transfer technique[41]. We first identified a thick hBN layer (40nm) and then used it to stack the top two $WS_2$ monolayers followed by stacking the tunnel region. The final stack structure from top to bottom is hBN / $WS_2$ / hBN / $WS_2$ / hBN / graphene / hBN / $WS_2$ / hBN / graphene / hBN.  There are 11 separate layers and the stacking was done continuously from top to bottom. Several stacking images are shown in Supplementary Section S2. The entire stack of van der Waal heterostructure was then transferred onto the DBR at temperature 120 ℃. Alignment was carefully done to make sure each graphene flake sits exactly on top of corresponding gold contact pad. After the transfer, the entire structure was soaked in chloroform for 2 hours to remove PPC residue followed by PMMA (495 A4 from Michrochem) spin coating to form a 200 nm top spacer layer. The final silver (40 nm) was deposited via e-beam evaporation for the top mirror of the microcavity. Details of each layer thickness and cavity response can be found in Supplementary Section S1.

***Optical measurement*** Angle resolved spectra were recorded using a homemade setup comprising of white light (broad band halogen source for reflection) and laser (PL measurement). The setup is coupled with Princeton Instruments monochromator with a PIXIS: 256 EMCCD camera. A 100X, 0.7 NA objective was used for all measurements. The polariton dispersion is revealed by imaging the back aperture of the microscope objective (Fourier plane) on to the camera. All measurements were done at room temperature.

**Data Availability:** Data are available on request from the authors

**Acknowledgements:** We acknowledge support from the National Science Foundation through the EFRI-2DARE program (EFMA-1542863), MRSEC program 420634 and the ARO MURI program (W911NF-17-1-0312). The authors also acknowledge the use of the Nanofabrication Facility at the CUNY Advanced Science Research Center for the fabrication of the devices.

**Author Contributions:**

V.M., J.G., B.C. conceived the experiments. J.G., B.C. M.K. fabricated the devices and performed the measurements. B.C., J.G., V.M. performed data analysis. All authors contributed to write the manuscript and discuss the results.

**Competing Interests:** The authors declare that they have no competing financial interests.

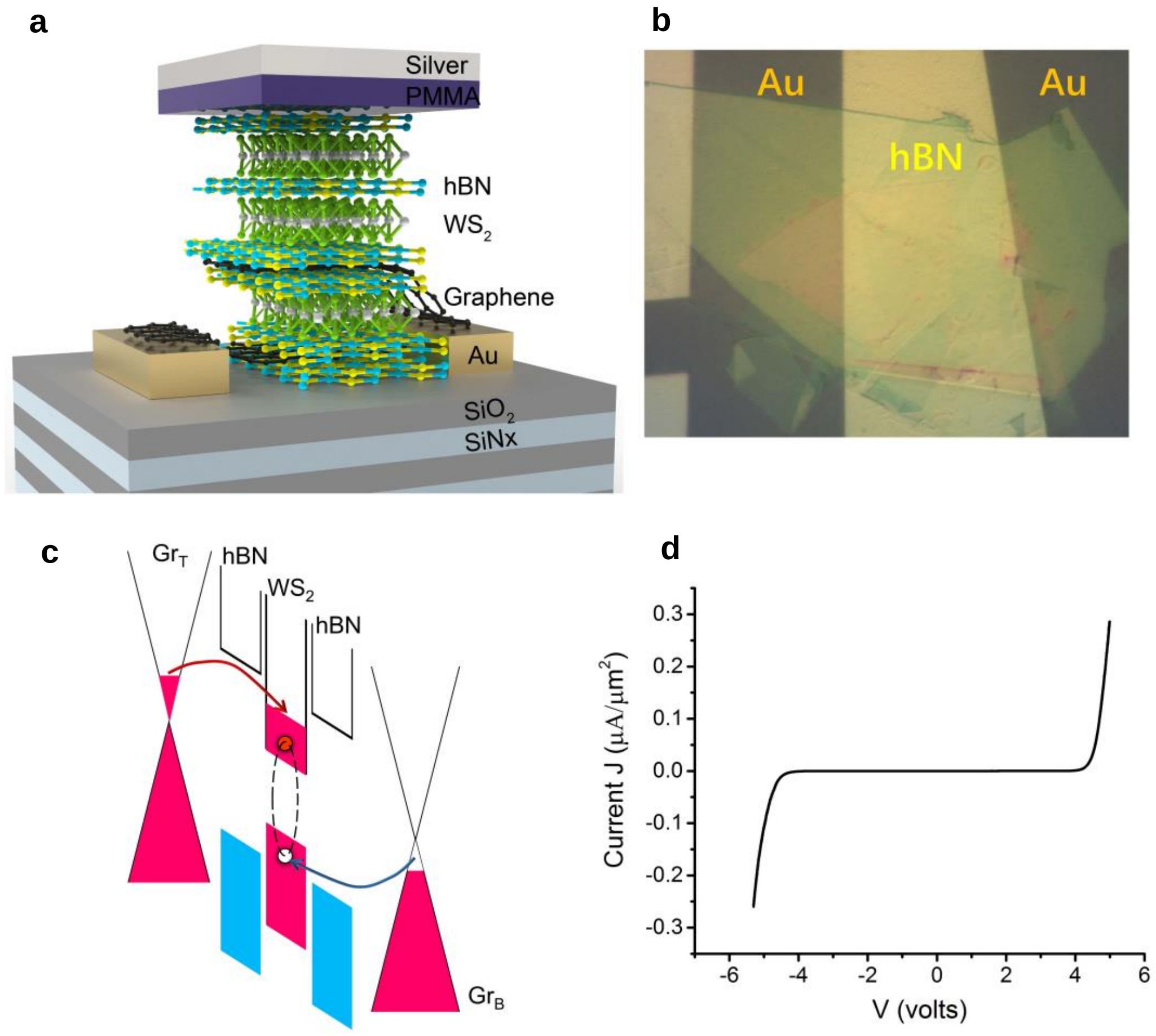

**Figure 1 | Device schematic and tunneling mechanism. a**, Schematic of the device. Thickness of each layer can be found in Supplementary Section S1. **b**, Optical image of the stacking before the top cavity is grown. Gold contacts and top hBN are labeled. Due to the large reflection from DBR substrate, only top hBN is observable. More images of sample fabrication is shown in Supplementary Section S2. **c**, Band diagrams at high bias above threshold. Electron (hole) can tunnel through hBN into $WS_2$ conduction (valence) band. Top (bottom) graphene is labeled as $Gr_T$ ($Gr_B$). **d**, Tunneling current as a function of bias voltage.

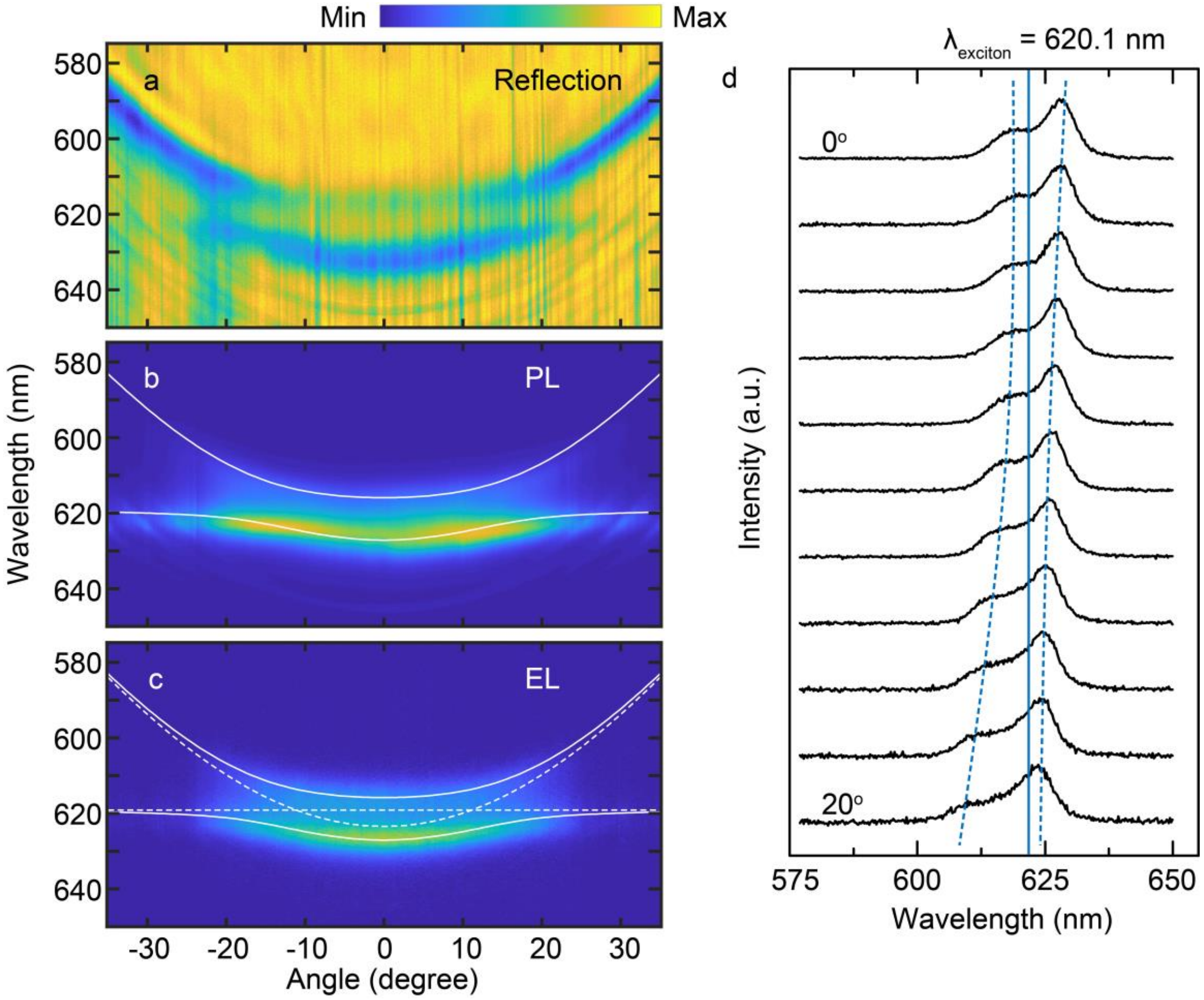

**Figure 2 | Polariton dispersion. a**, Angle resolved reflectance and **b** photoluminescence. White solid lines in **b** are the guide lines following the polariton branches. **c**, Angle resolved electroluminescence at current injection of 0.10 $\mu$A/$\mu$m$^2$. The solid lines are coupled oscillator model fit to the polariton branches. The dashed lines are bare cavity and exciton resonances. The extracted Rabi splitting and detuning are 33 meV and -13 meV, respectively. **d**, EL spectral plots at different angles from sectional slice in **c**. The (blue) solid vertical line indicates exciton emission wavelength ($\lambda_{exciton}$) at 620.1 nm. Dashed trend lines indicate upper and lower polariton branches.

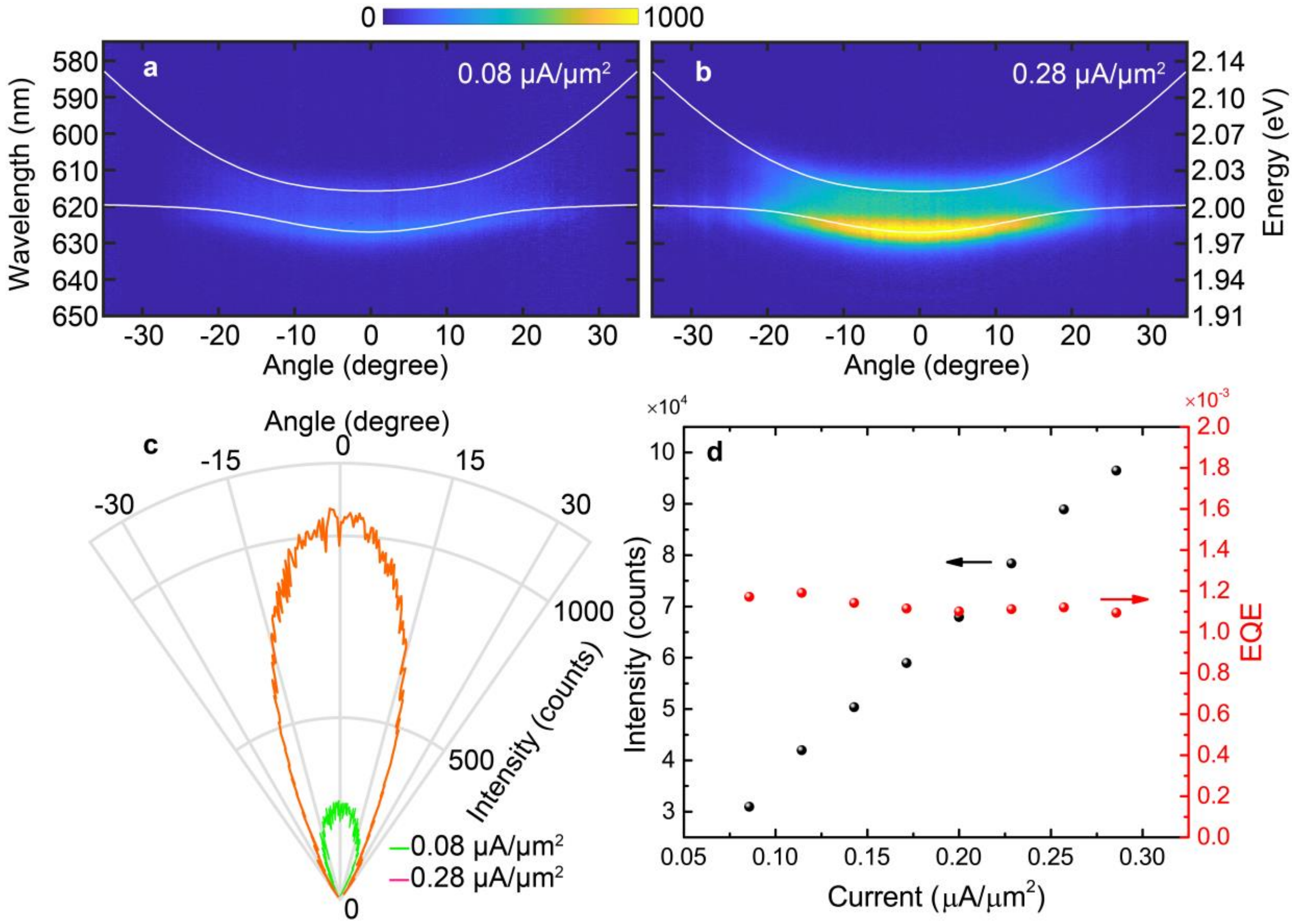

**Figure 3 | Current dependent polariton EL. a**, Polariton dispersion from EL under current injection of 0.08 $\mu$A/$\mu$m$^2$ and **b** 0.28 $\mu$A/$\mu$m$^2$. White solid lines are the guide lines following the polariton branches. **c**, Polar plot from different current density. The emission angular distribution pattern does not change within the range of applied current. **d**, Integrated EL intensity (black) and EQE (red) as a function of current density. The EL process is in the linear regime within the range of current applied.

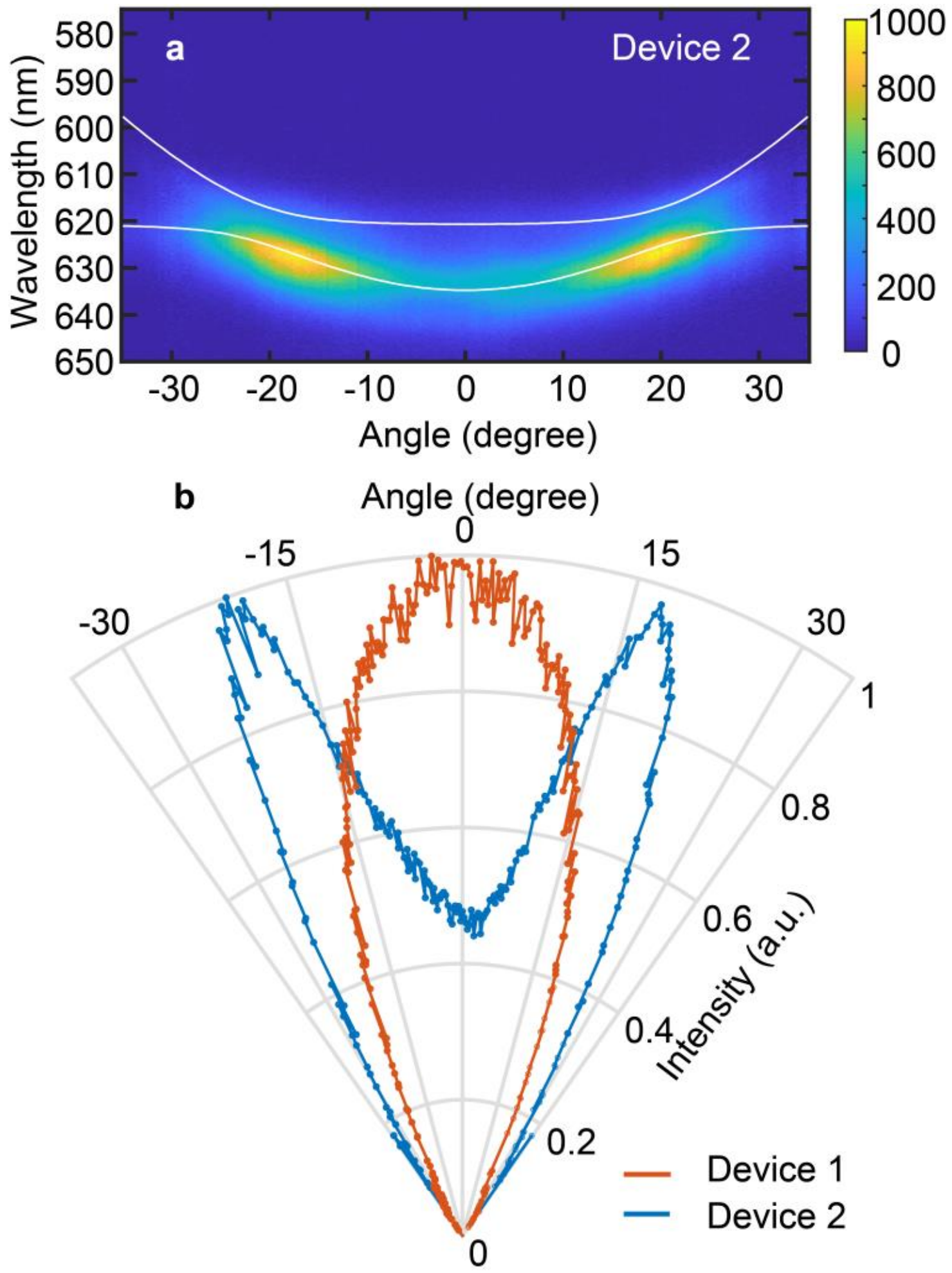

**Figure 4 | Detuning dependence on EL.** **a**, Device 2 angle resolved EL (0.2 $\mu$A/$\mu$m$^2$) with a large negative detuning (-43 meV). A strong bottle neck effect was observed. **b**, Normalized angular emission pattern from device 1 (Fig. 2c) and device 2 (Fig. 4a) showing distinct emission patterns owing to the difference in detuning.

## Table of Contents

# S1. Cavity mode distribution

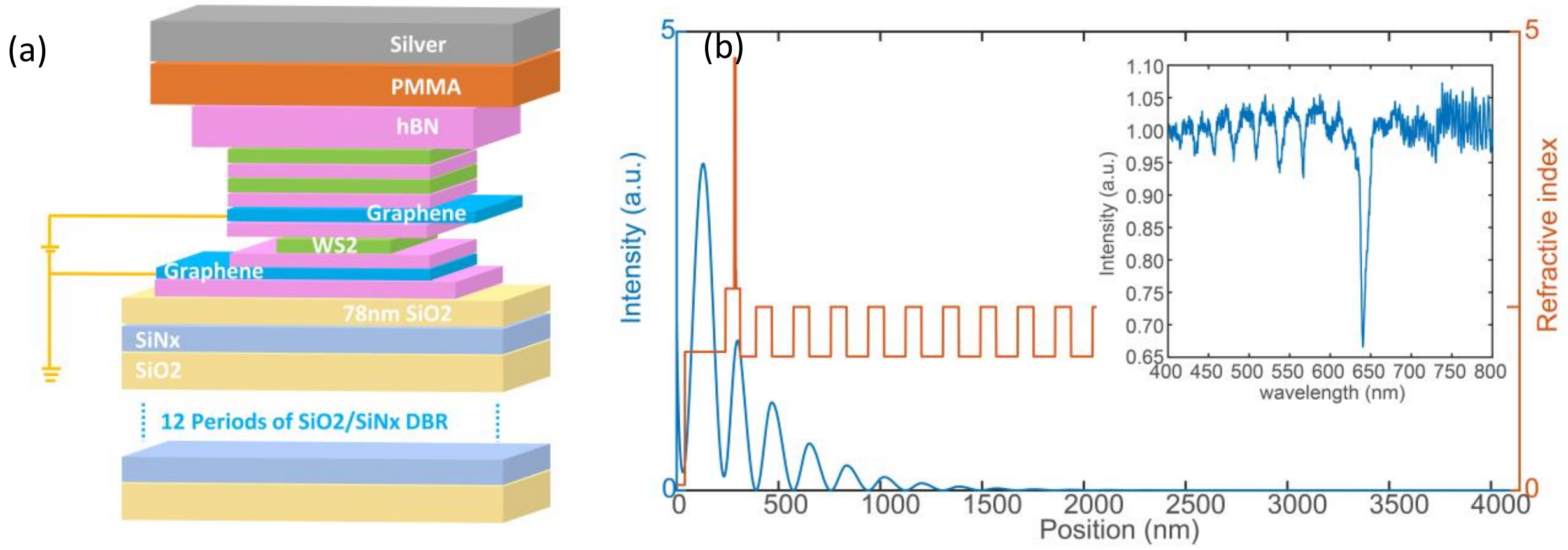

**Figure S1 | Device schematic and cavity resonance. a**, Device schematic. **b**, Mode distribution for wavelength at 620.1nm which is the $WS_2$ exciton energy. Position '0' relates to the top surface which is silver. Thickness of the cavity is tuned to have the mode profile's second maximum position falling on the $WS_2$ region. Inset shows the bare cavity mode with 8.2nm (24meV) linewidth.

| **Material** | **Silver** | **PMMA** | **hBN** | **$WS_2$** | **hBN** | **$WS_2$** | **hBN** | **Graphene** | **hBN** | **$WS_2$** | **hBN** | **Graphene** | **hBN** | **$SiO_2$** | **DBR-SiNx** | **DBR-$SiO_2$** |
|---|---|---|---|---|---|---|---|---|---|---|---|---|---|---|---|---|
| **Thickness (nm)** | **40** | **200** | **40** | **0.7** | **2** | **0.7** | **3** | **0.4** | **2** | **0.7** | **2** | **0.4** | **20** | **78** | **77.5** | **106.2** |

**Table S1 | Thickness of each layer in Figure S1a.**

## S2. Stacking images

This section shows the optical images of stacking steps after each $WS_2$ layer was picked up.

| Stacking order | Image on PPC film | $WS_2$ image |
|---|---|---|
| 1. First $WS_2$ 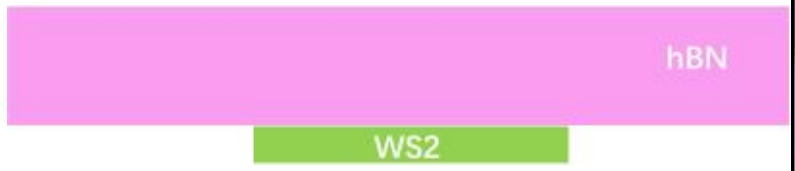  | 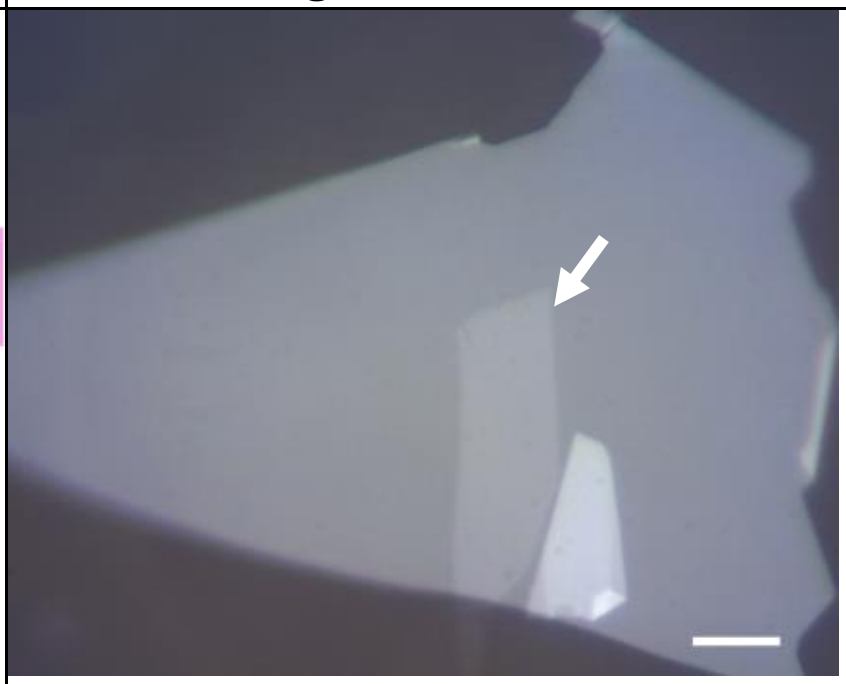 | 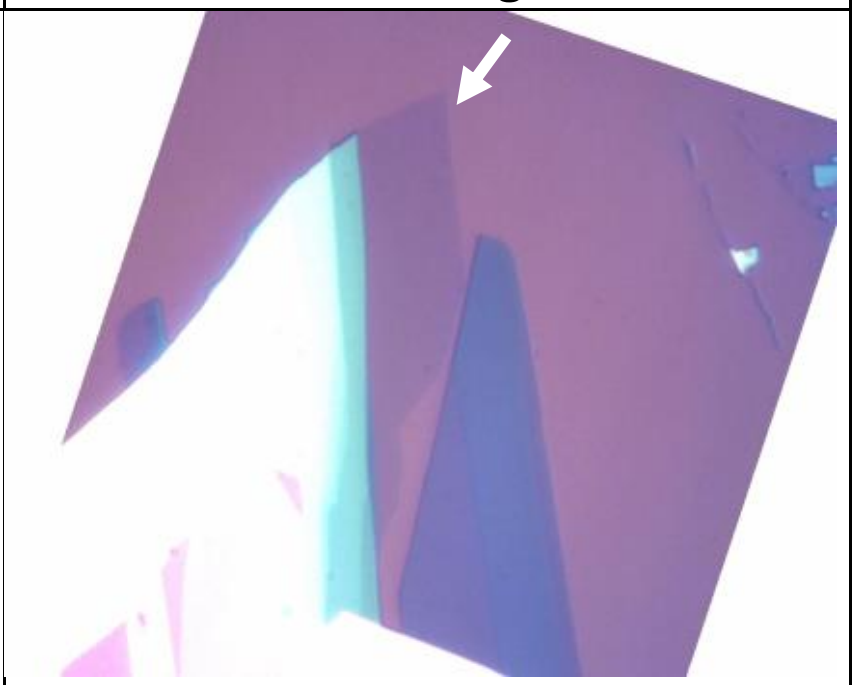 |
| 2. Second $WS_2$ 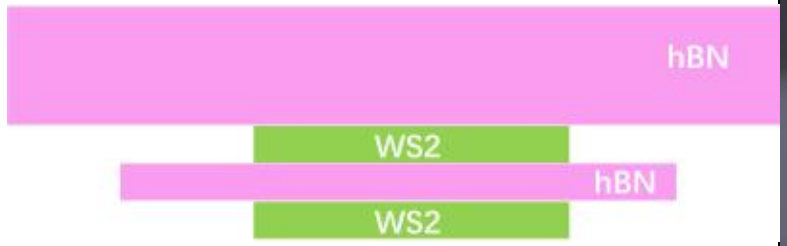  | 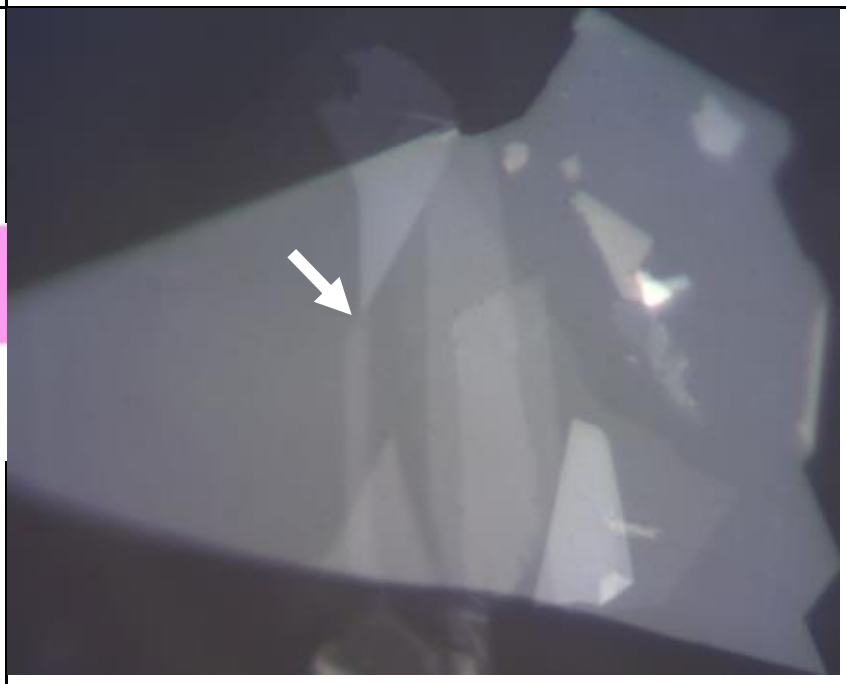 | 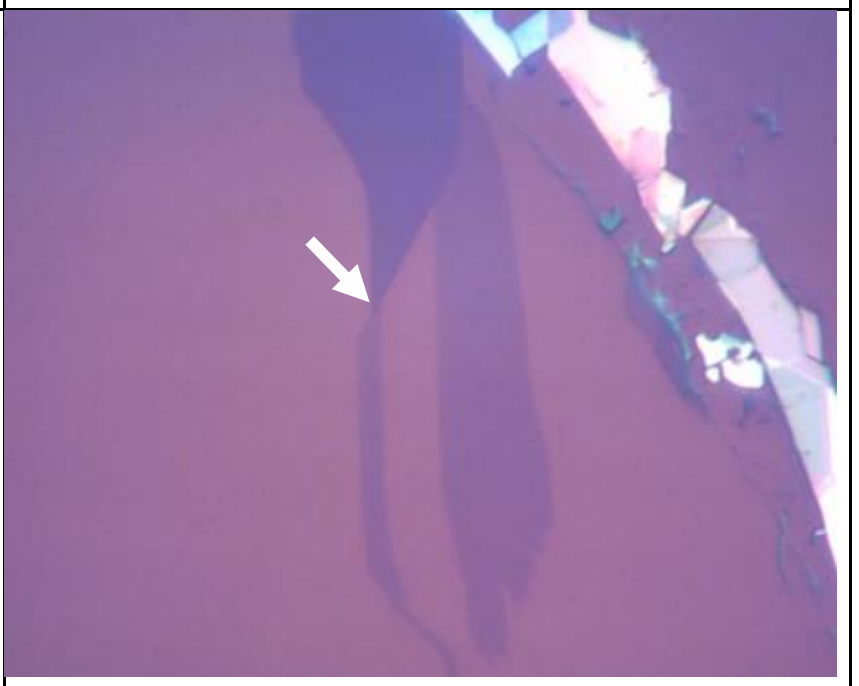 |
| 3. Third $WS_2$ 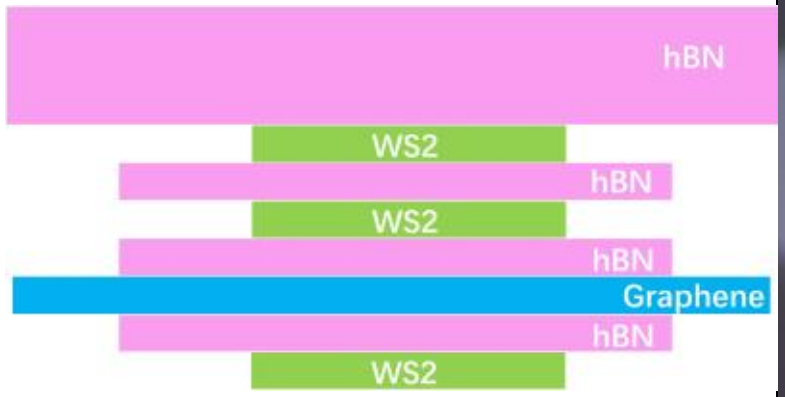  | 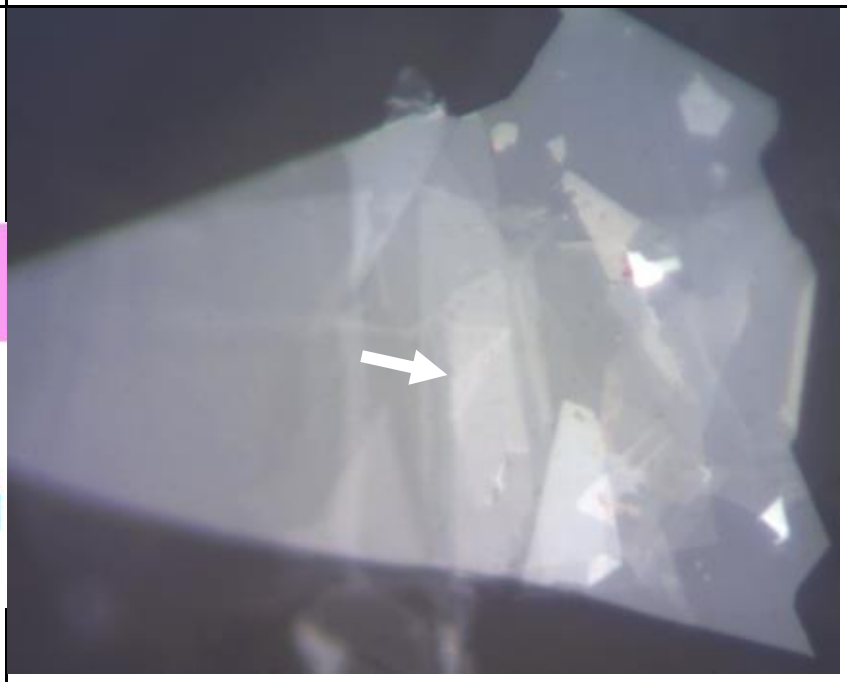 | 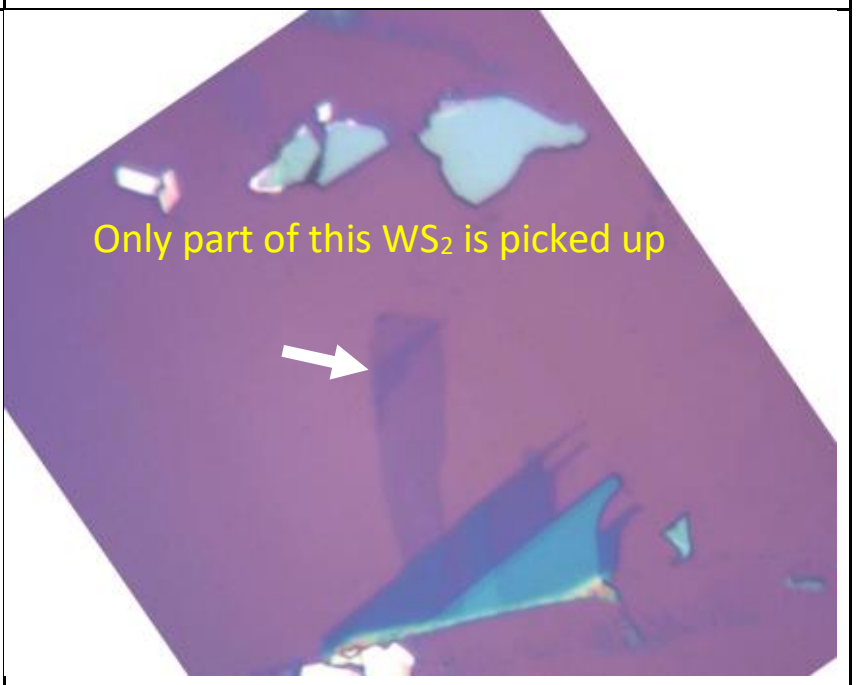  |

**Table S2 | Optical images of the stacking at different steps. Left:** Schematics of the stacked layer order. **Middle:** Optical image of the PPC film at different steps. **Right:** Microscope optical image of different $WS_2$ layers. In each row, there are two white arrows in middle and right columns pointing at the same position for the guiding of view to ensure the $WS_2$ has been successfully picked up at each step. Monolayer $WS_2$ is easy to identify due to the color contrast, for example, in the first $WS_2$ image, the white arrow is pointing at the monolayer region. All the figures have the same scale. A 10 $\mu$m scale bar is shown in the first row PPC film image.

## S3. EL spatial image

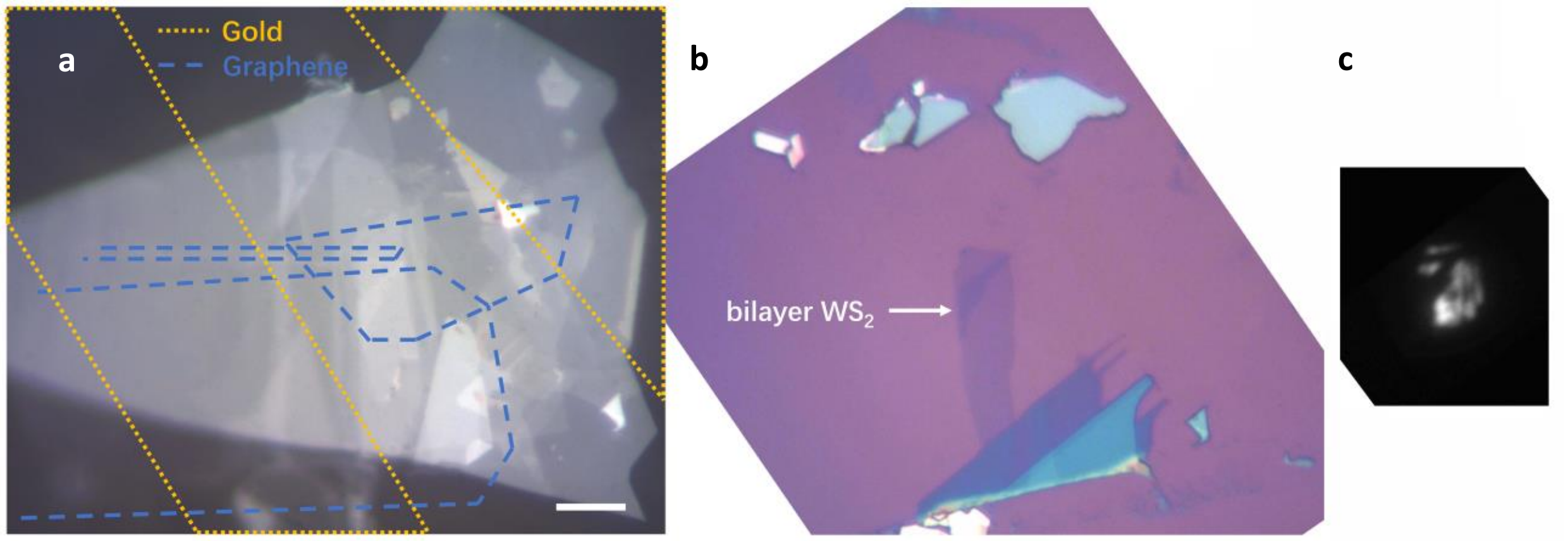

**Figure S3 | EL spatial image. a**, Sample optical image on the PPC film after the third $WS_2$ layer was picked up. The graphene (long dash) and gold contacts (short dash) outlines are labeled with different colors. The scale bar is 10$\mu$m. All the images have the same scale bar. **b**, Optical image of the third $WS_2$ layer on $SiO_2$/Si substrate before the picking up. There is a small bilayer region shown by the white arrow. **c**, EL spacial image at 0.28 $\mu A/\mu m^2$. Because only the third $WS_2$ layer is inside the tunneling region, as shown in **Table S2**, row 3, when a bias is applied across the graphene contacts, only the third layer $WS_2$ will show EL. The bilayer region in the third $WS_2$ shows much less EL due to its indirect band gap.

## S4. Device 2 data

We made another device by **first** stacking and transferring the tunneling structure, which has only one $WS_2$ monolayer inside, followed by stacking and transferring another top two $WS_2$ monolayers (hBN/$WS_2$/hBN/$WS_2$), unlike the device 1 stacking steps, which stacked the top two $WS_2$ layers first and then stacked the tunneling structure. The final structure of device 2 is the same as device 1 as shown in Fig. S1a. The only difference is that the third hBN (count from top to bottom) layer thickness is 30 nm because this hBN layer acts as the first picking up layer for the tunneling structure in device 2. The rest of PMMA and $SiO_2$ thickness were also tuned to match to mode position. One of the advantages of trying this method is that, by separating the stacking process into two (one has 7 stacking layers, another has 4 stacking layers), the device fabrication will have a much higher success rate. The original stacking encounters continuous 11 layers with much higher chance of damaging the thin graphene contact by more stacking-picking up process. Images of the first stacking process which evolves 7 stacking layers are shown in **Table S4** (below). Device 2 also shows strong coupling with Rabi splitting 27meV.

<table>
<tr><th>Stacking order</th><th>Stacking image</th><th>Current stacked layer image</th></tr>
<tr><td>1.<br>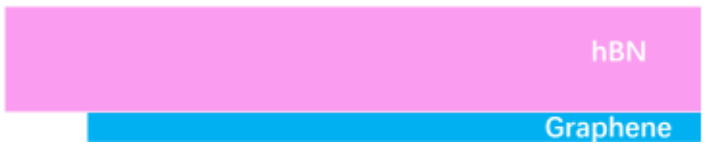<br></td><td>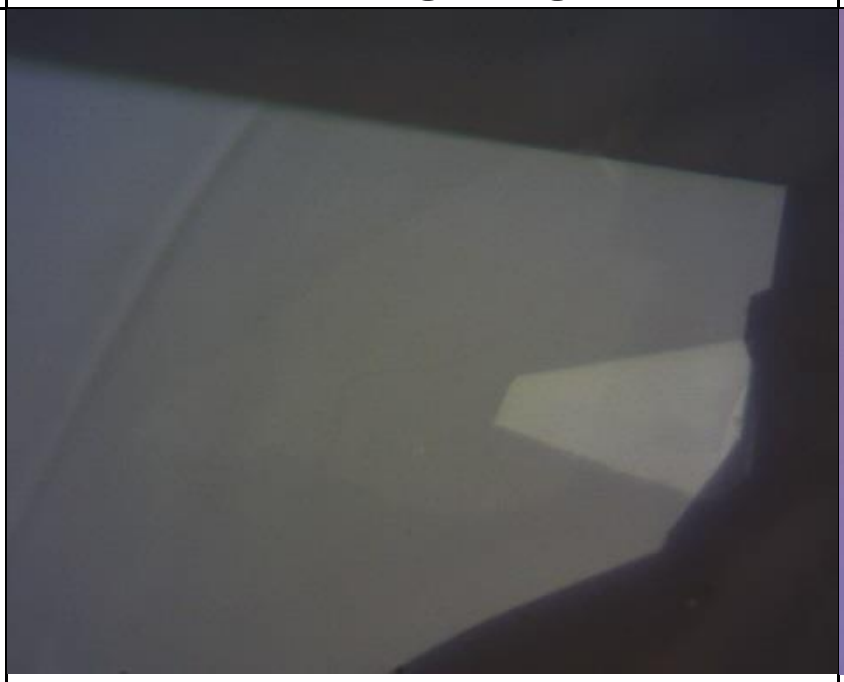</td><td>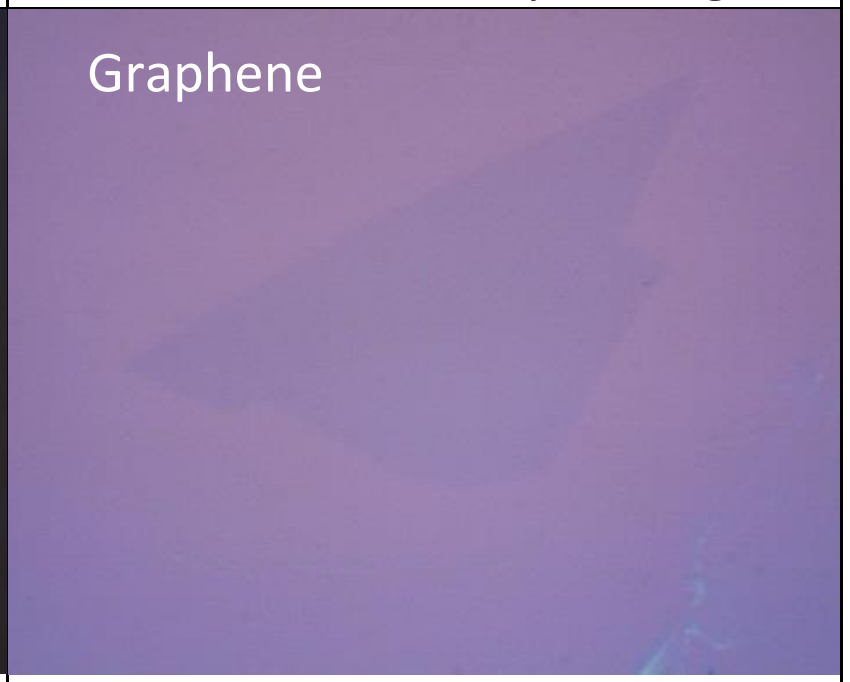<br></td></tr>
<tr><td>2.<br>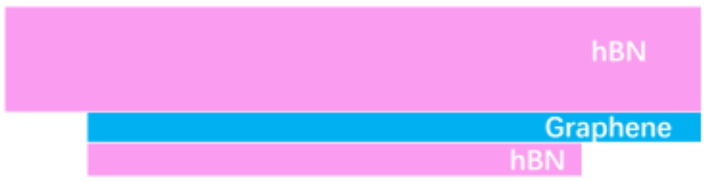<br></td><td>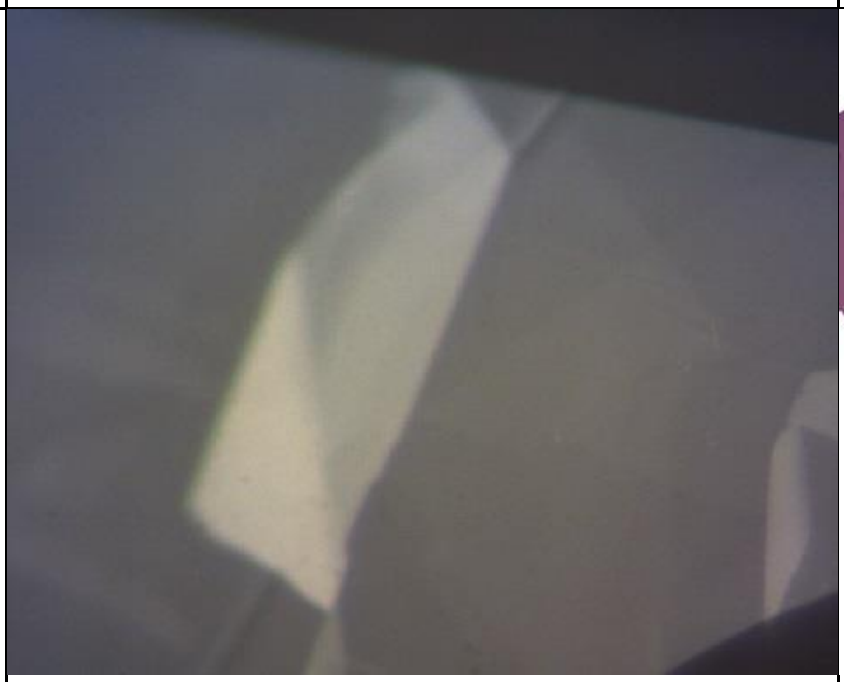</td><td>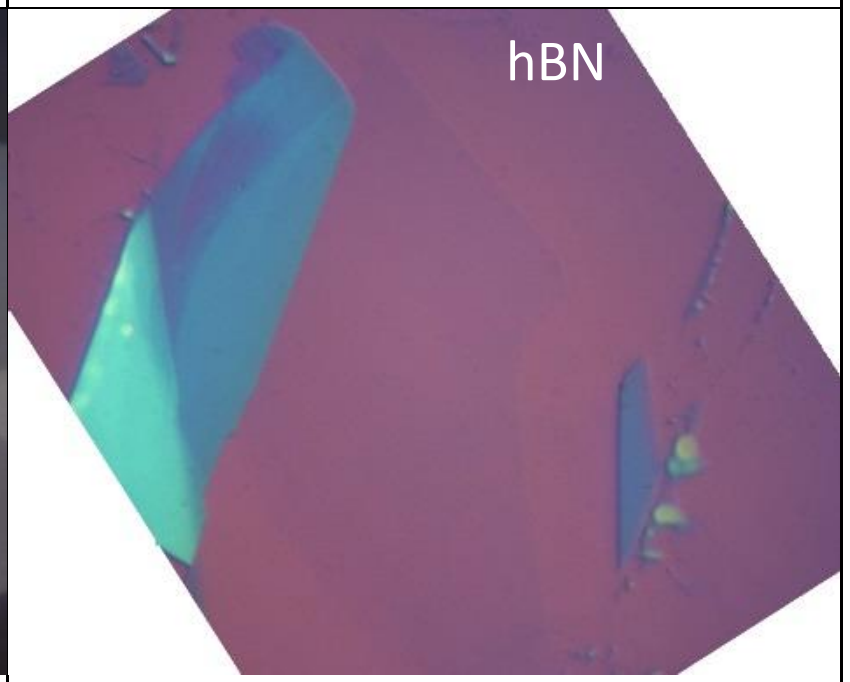<br></td></tr>
</table>

<table>
<tr>
<td>3.<br>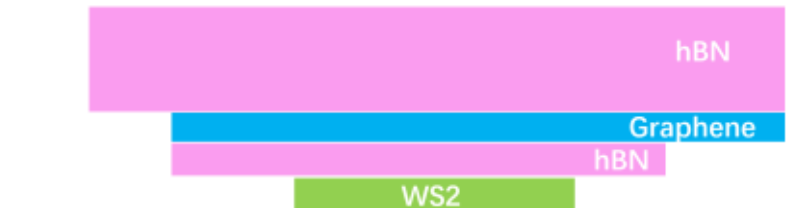</td>
<td>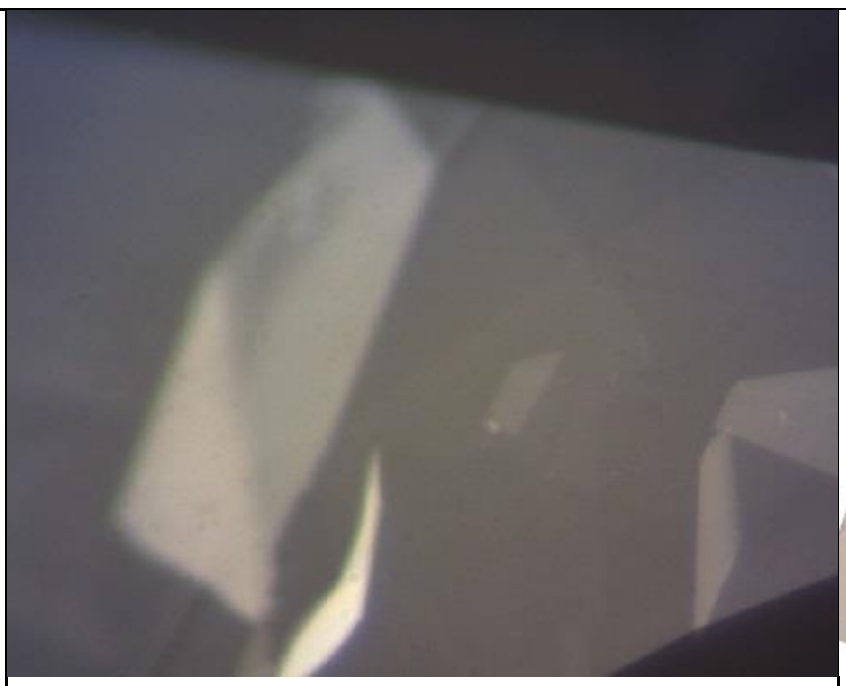</td>
<td>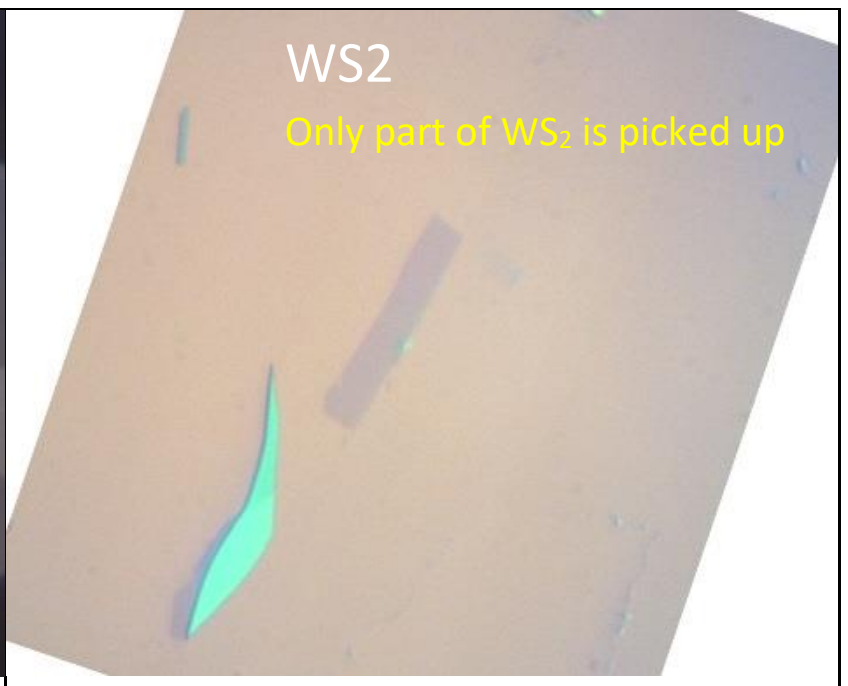</td>
</tr>
<tr>
<td>4.<br>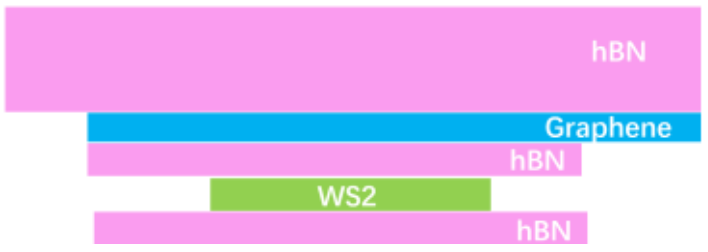</td>
<td>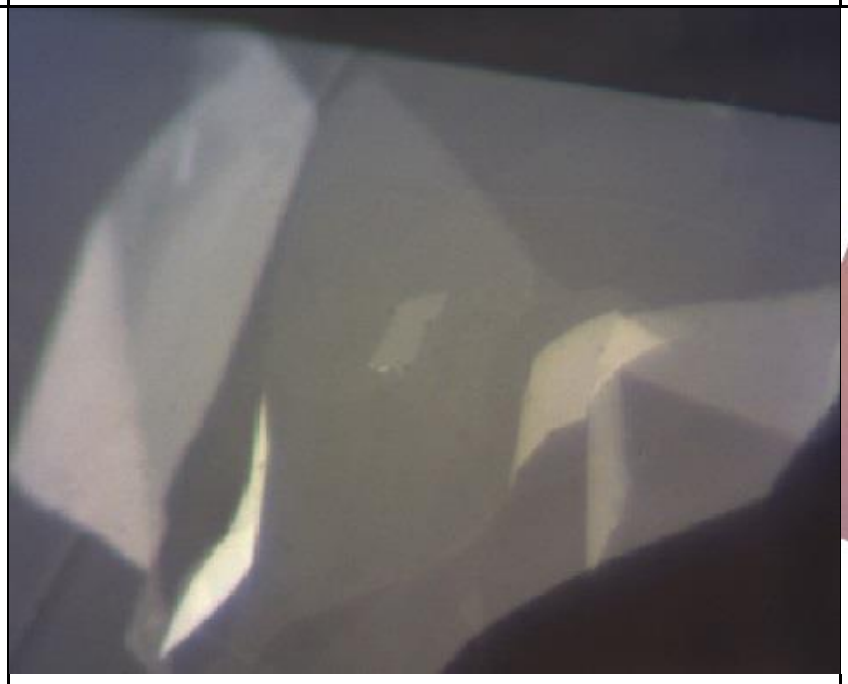</td>
<td>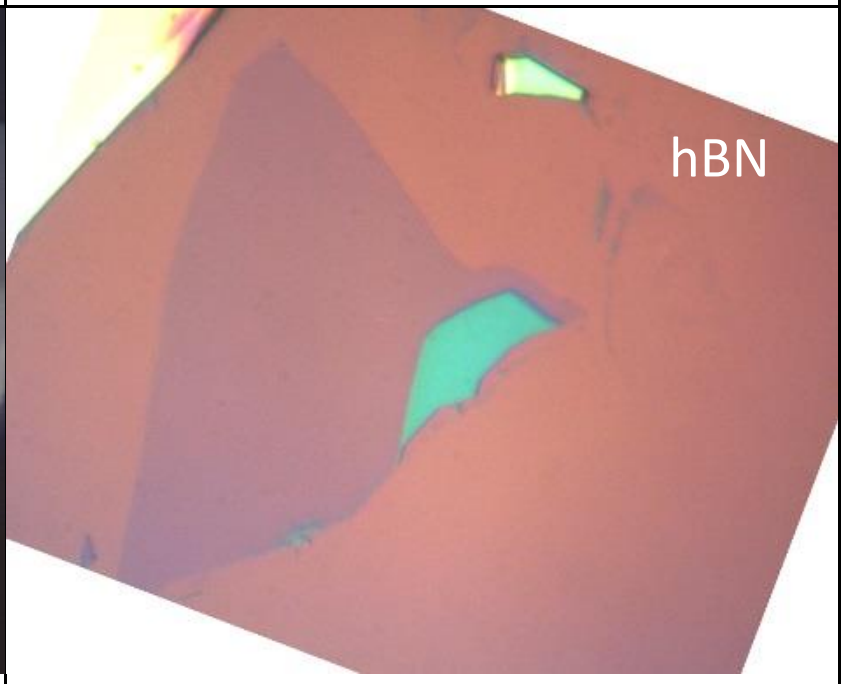</td>
</tr>
<tr>
<td>5.<br>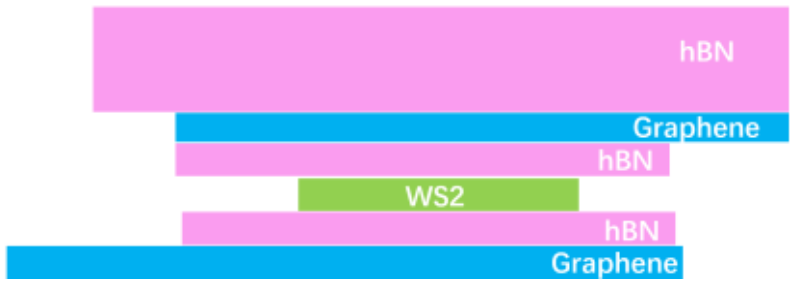</td>
<td>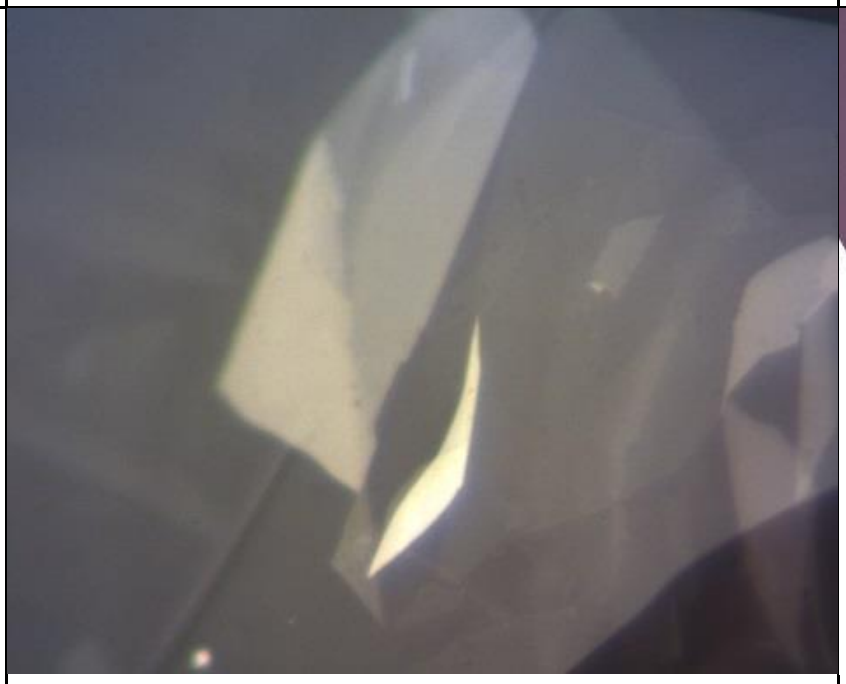</td>
<td>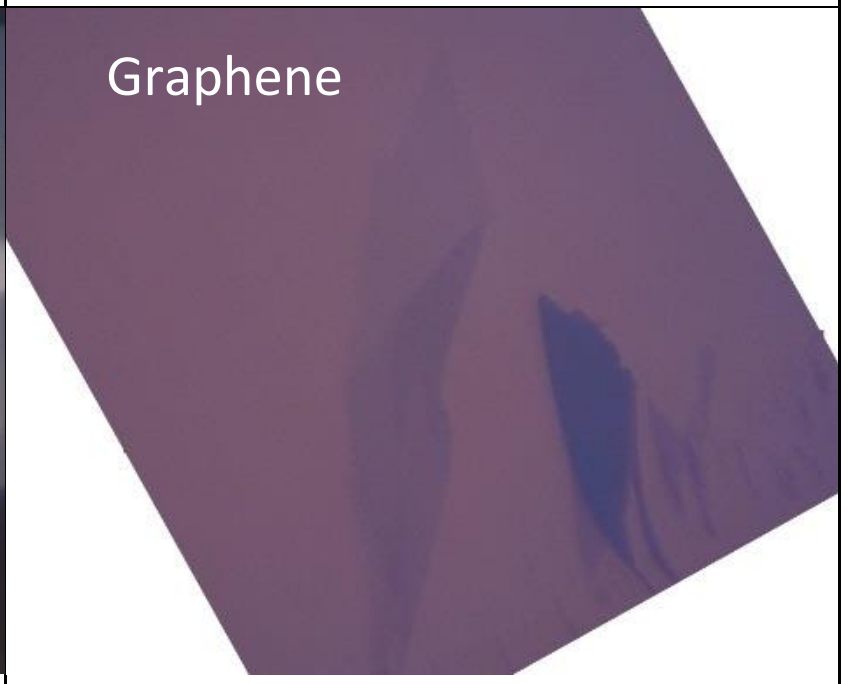</td>
</tr>
<tr>
<td>6.<br>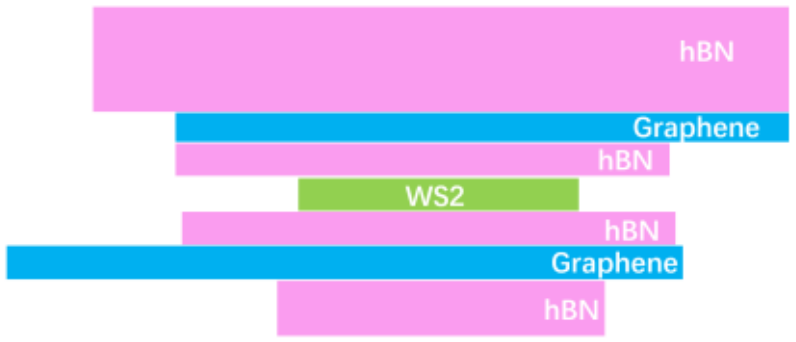</td>
<td>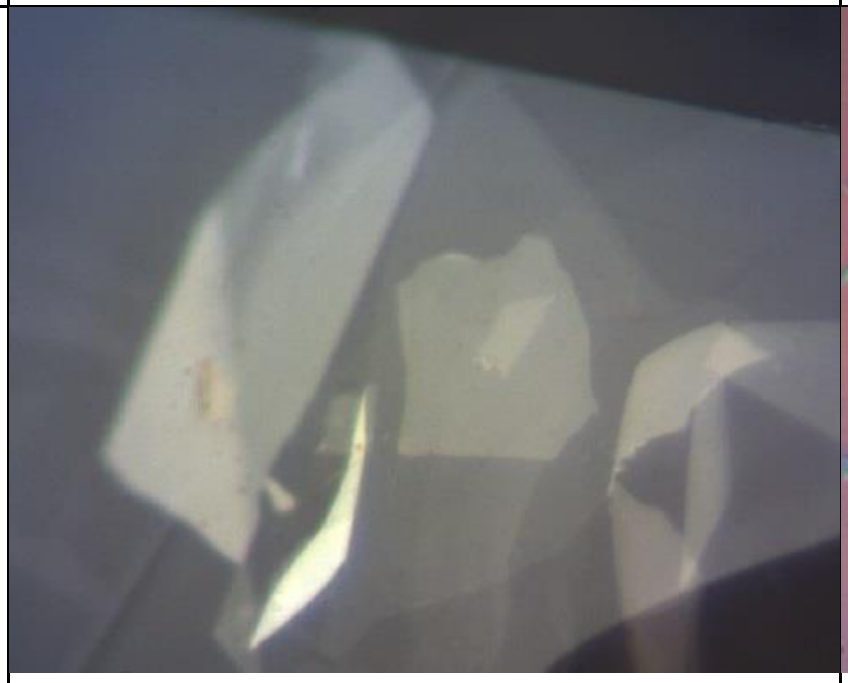</td>
<td>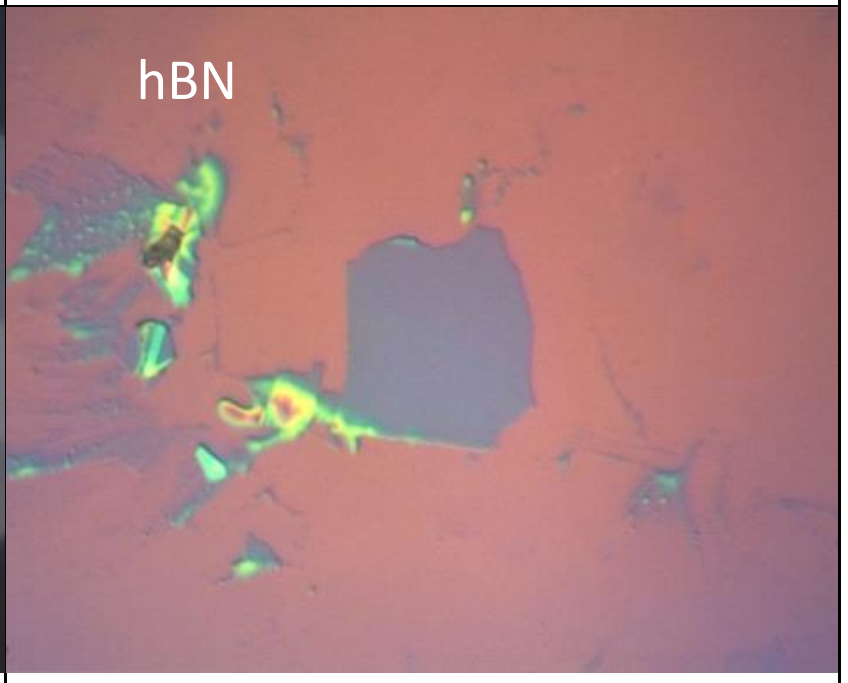</td>
</tr>
</table>

**Table S4 | Device 2 stacking images.**

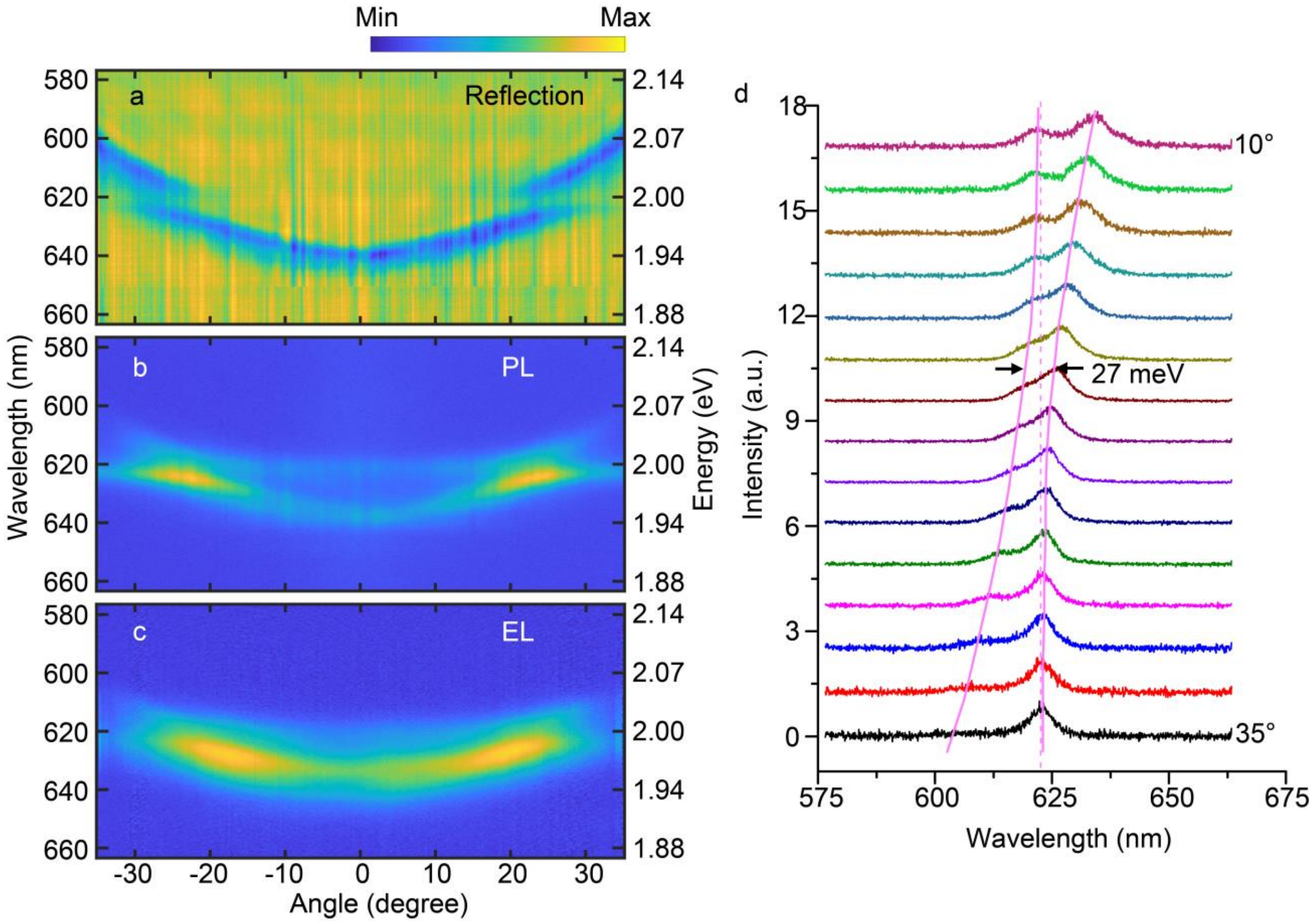

**Figure S4 | Angle resolved spectra from device 2.** Angle resolved reflection (**a**) PL (**b**) EL (**c**) spectra. **d**, EL linecut from 10 degree to 35 degree showing 27 meV Rabi splitting.

## S5. External Quantum efficiency

The external quantum efficiency (EQE) is defined as the number of photon emitted per tunneling carrier Ne/i (N is the photon emitted every second from the total tunneling monolayer $WS_2$ area, e is the electron charge, i the current passing through the total tunneling monolayer $WS_2$ area). N is defined as

$$\mathrm{N} = \frac{N_{detected}}{\alpha_{total}}$$

$N_{detected}$ is the intensity detected by the detector per second from $WS_2$ tunneling emission. $\alpha_{total}$ is the total collection efficiency of the whole system. Light emitted from monolayer is collected by an objective with N.A 0.7 and guided through mirrors and lenses into the detector. The detector is set to a high gain factor. All those will affect the actual counts read by the detector. In order to know N, we need to measure $\alpha_{total}$ for a given set up that is exactly the same as the one we use to detect the tunneling emission. We did this by shining a laser that has the same wavelength (620.1nm) as our tunneling emission to a perfect reflector and the reflected beam was guided along the same path as the tunneling emission into the detector. The laser power on the reflection standard surface is measured to be $P$. The measured laser counts is $N_{laser}$ per second. $\alpha_{total}$ is calculated as:

$$\alpha_{total} = \frac{N_{laser}}{P} * E$$

E is the laser energy which is 2 eV.

So N is:

$$\mathrm{N} = \frac{N_{detected}}{N_{laser} * E} * P$$